\documentclass[a4paper,12pt]{article}
\usepackage{amsmath,amsfonts,amssymb}
\usepackage{cite}

\begin{document}

\begin{center}
{\Large\bf On the hypothesis of a second Higgs boson near 0.5 TeV}
\end{center}
\bigskip
\begin{center}
{S.S. Afonin
}
\end{center}

\begin{center}
{\footnotesize\it 
Saint Petersburg State University, Universitetskaya nab. 7/9,
St. Petersburg, 199034, Russia
}
\end{center}

\bigskip

\begin{abstract}
Recently, a second Higgs-like boson $h'$ with the mass near 0.5~TeV was predicted from a dual holographic model
describing the hypothetical strongly-coupled sector beyond the Standard Model. We analyze under what conditions this prediction can be reproduced within
the framework of more traditional models for describing a strongly-coupled field theory ---  the spectral sum rules and the Nambu--Jona-Lasinio model
in the scalar channel. 
It is shown that the results of both approaches are close and lead to this prediction if the covariant four-momentum cutoff in them is identified with the unitarity bound on the Higgs boson mass, and also under the assumption that the strongly coupled sector beyond the Standard Model is described by some quantum field theory based on the $SU(2)$ gauge group.
We also present additional arguments suggesting that a mass of about 0.5 TeV would be natural for a heavy analogue of the Higgs boson,
if it exists.
\end{abstract}


\section{Introduction}

The origin of the Higgs potential in the Standard Model (SM) of particle physics remains a mystery.
It is widely believed that some physics beyond the SM (BSM) should be responsible for the known phenomenological form of the SM Higgs sector.
Many models for a hypothetical BSM physics predict new fundamental particles. The simplest option to extend the SM Higgs sector
consists in adding a singlet Higgs-like scalar boson. Such a minimal extension of the SM has various attractive features and
is potentially able to solve several fundamental problems (see, e.g., the relevant discussions and references
in~\cite{MartinLozano:2015vtq,Godunov:2015nea,Falkowski:2015iwa}).
The given scenario entails an important question about the mass of this second scalar boson since the value of its mass is
crucial for the experimental searches for this state.

We have addressed this question in our recent paper~\cite{plb}, where a dynamical model was proposed that predicted the second scalar
boson near 0.5~TeV. More accurately, for the fixed Higgs mass $m_h=125$~GeV, the prediction was $m_{h'}\simeq515$~GeV.
The authors of recent Ref.~\cite{Asiain:2023myt} named this putative state $h'(515)$, among other BSM candidates for the second scalar boson.
The $h'(515)$ was obtained in a particular model built within the framework of a specific
phenomenological approach. This, of course, does not look very convincing. On the other hand, it is well known that if some theoretical
result in physics is true then this result can be usually obtained in several different ways. Following this principle,
in the present work we will consider the question of whether the $h'(515)$ can be reproduced in a natural way within the framework
of other, more traditional approaches which have been exploited in the BSM  phenomenology.

First we should briefly recapitulate the main ideas underlying our prediction in Ref.~\cite{plb}. A popular direction of searches for BSM
physics is based on a general conjecture that this physics arises from a strongly-interacting dynamics described by some field
theory which may be similar to QCD in the strongly-coupled regime and that the Higgs boson represents a composite state
emerging from this dynamics~\cite{Bellazzini:2014yua}. In addition, it is often assumed that the SM Higgs particle is a pseudogoldstone boson in
this dynamics by analogy with the pion in QCD (see~\cite{Contino:2010rs} for a review). The given concept suggests that various phenomenological
approaches which have been traditionally used for description of low-energy QCD may be applied to the hypothetical strongly-interacting
BSM sector. As a consequence, predictions of extra composite states (some hadrons composed of quarks) given by those approaches can be
then converted into predictions of new BSM particles. A prominent example of this strategy is the use of five-dimensional
holographic approach, to a large extent borrowed from the holographic models for QCD, in modeling the Higgs sector of BSM physics~\cite{Contino:2010rs}.
Following this line of thinking, we applied the bottom-up holographic model of Ref.~\cite{Afonin:2021zdu}, originally constructed for description
of Regge spectroscopy of light mesons, to the hypothetical BSM sector. The obtained numerical relation between masses of Higgs-like
scalars, $m_{h'}\approx 4.12m_h$~\cite{plb}, turned out to be close to the approximate relation between masses of the lightest non-strange
scalar mesons, $m_\sigma\simeq 4m_\pi$, where the $\sigma$-meson represents the scalar isoscalar resonance $f_0(500)$ in the
modern Particle Data~\cite{pdg}.

The first phenomenological approach, which we will consider in the next Section, is based on the borelized spectral sum rules in QCD (ITEP sum rules)
(see, e.g., the reviews~\cite{Colangelo:2000dp,Gubler:2018ctz}).
This method was formulated and developed by Shifman, Vainshtein and Zakharov in the late 70th~\cite{svz1,svz2} and became extremely popular in the hadron
phenomenology in the 80th since it was often revealing a remarkable agreement with the experimental data.
Roughly speaking, the given method assumes that a quark current (that interpolates a hadron --- a quark-antiquark pair in
the simplest case) created in the vacuum medium of strongly interacting quarks and gluons does not perturb this vacuum in the first approximation.
The unknown non-perturbative vacuum can then be parameterized by a set of its universal phenomenological characteristics --- the vacuum condensates.
The quark currents with different quantum numbers react differently to the vacuum medium and, as a consequence, the hadrons coupled to these
currents will have different physical characteristics --- masses, magnetic moments, decay constants, and formfactors. It turned out that these
characteristics can be calculated (with a typical accuracy to about 10\% -- 20\%) from a semi-analytical machinery of spectral sum rules in QCD which relies on
the analyticity of the correlators (two-point functions) of interpolating currents and asymptotic freedom of QCD. The former allows one to derive
dispersion relations which relate the integrated spectral function (the imaginary part of correlator) in the deep Euclidean region with resonances
in the physical region. The latter is exploited to systematically compute the correlator in the Euclidean region using the Operator Product Expansion
(OPE) supplemented by condensates (the phenomenological v.e.v. of QCD operators which replace the operators in the original wilsonian OPE). The given
method was rather successful not only in describing hadronic ground states but also in extracting non-observable QCD parameters --- the values of
quark masses and QCD coupling at various energy scales~\cite{Colangelo:2000dp,Gubler:2018ctz}.
The QCD sum rule predictions proved their reliability in most cases when one needs to determine an unknown hadronic parameter.

The second approach we will consider (Section~3) is based on a well known similarity between descriptions of the Higgs boson in the SM and the
$\sigma$-meson in effective models for low-energy strong interactions (see, e.g., the review~\cite{Schumacher:2014jga}). By virtue of this similarity,
the $\sigma$-meson is often called ``the Higgs boson of strong interactions'' for its role in model descriptions of origin of
nucleon mass~\cite{Schumacher:2014jga}. A fruitful microscopic model for composite $\sigma$-meson as a quark-antiquark state appeared within the framework of the
Nambu--Jona-Lasinio (NJL) model (see, e.g., the classical review~\cite{klevansky}). The standard NJL model approximates the low-energy strong
interactions by local four-fermion interactions. In this regard, the NJL model in theory of strong interactions bears resemblance
to the Fermi model in theory of weak interactions. This resemblance suggests that integrating out the gauge boson degrees of freedom
in a gauge theory one obtains, in the first low-energy approximation, an effective theory of local four-fermion interactions regardless of
whether this gauge theory was in the higgs or in the confinement phase. An extension of this phenomenologically working observation to a
hypothetical BSM gauge theory looks thus reasonable. And indeed, various NJL-type models have been widely used in the past as an approximation to
some supposed new BSM strong dynamics (see, e.g.,~\cite{Hill:2002ap} and references therein). Interestingly, the SM for any fermion can always be
rewritten as a kind of some generalized NJL model~\cite{Hasenfratz:1991it}. Perhaps the most frequent use of the NJL model for this purpose was within the
framework of top quark condensation~\cite{Cvetic:1997eb}. This old idea relies on the assumption that the underlying physics above
a compositeness scale leads to effective four-quark interactions strong enough to induce quark-antiquark condensation into composite Higgs
field(s). The minimal $t\bar{t}$ condensation framework, however, was ruled out but it gave rise to many extensions of the minimal framework~\cite{Cvetic:1997eb}.

In the discussions of Section~4 we provide a couple of heuristic arguments indirectly confirming the consistency of our result. 
Our main conclusions are collected in Section~5.

\section{BSM scalar boson from spectral sum rules}

As we mentioned in the Introduction, the spectral sum rules in QCD (namely their variety called the Shifman--Vainshtein--Zakharov or ITEP sum rules~\cite{svz1,svz2}) were
very popular semiquantitative non-perturbative method in the hadron physics, especially in the 80th\cite{Colangelo:2000dp,rry}. Actually they are still actively used
in various forms (see, e.g., some examples and many references of the their use in the borelized form in~\cite{AlamKhan:2023ili},
in moments form in~\cite{AlamKhan:2023kgs} and in the
planar limit of QCD in~\cite{jhep,Afonin:2016evz}). We briefly recapitulate the main idea of this method in the area of extraction of ground state masses
of light hadrons, where the borelized variant is exploited.

One considers the two-point correlation function of some hadron current $j(x)$,
\begin{equation}
\label{12}
\Pi(q^2)=i\int d^4x\, e^{iqx}\left\langle0|\text{T}\left\{j(x),j(0)\right\} \right|0\rangle,
\end{equation}
and calculates the OPE for~\eqref{12} in the Euclidean space ($q^2=-Q^2$) in the following schematic form,
\begin{equation}
\label{13}
\Pi(Q^2)=a_0Q^2\ln{\frac{Q^2}{\mu^2}} + \sum_{n=1}^\infty a_n\frac{C_n}{Q^{2n}}.
\end{equation}
Here $C_n$ are phenomenological non-perturbative condensates (v.e.v. of various gauge-invariant QCD operators of canonical dimension $2n$
constructed from the quark and gluon fields; if there are several such operators, $i=1,2,\dots,$ then the summation over $i$, $a_n^iC_n^i$, is implied),
$a_n$ are perturbative coefficient functions depending on the normalization scale $\mu$, and the first logarithmic contribution (playing the role of
unit operator) arises thanks to the asymptotic freedom of QCD. Further one accepts the approximation in which the correlator~\eqref{12} is saturated
by one resonance (pole) term plus a perturbative continuum. This is motivated, first, by the observation that the heavier states are exponentially
suppressed in the Euclidean space and, second, by making use of the Borel transform of~\eqref{12},
\begin{equation}
\label{14}
L_M\Pi(Q^2)=\lim_{\substack{Q^2,n\rightarrow\infty\\Q^2/n=M^2}}\frac{1}{(n-1)!}(Q^2)^n\left(\frac{-d}{dQ^2}\right)^n\Pi(Q^2).
\end{equation}
This transform suppresses contribution of operators of dimension $2n$ by a factor of $1/(n-1)!$ and of heavier resonances (``radial
excitations'' with masses $m_n$) by a factor of $e^{-m_n^2/M^2}$, i.e., the Borel transform~\eqref{14} improves the convergence
and simultaneously enhances the relative contribution of ground state. Neglecting also the decay width, one writes the following
``one resonance plus continuum'' ansatz for the spectral function,
\begin{equation}
\label{15}
\text{Im}\Pi=f_h^2s\delta(s-m_h^2) + a_0 s\Theta(s-s_0),
\end{equation}
where $m_h$ is the mass of a hadron $h$ (that carries the quantum numbers of the current $j(x)$ in~\eqref{12}), $f_h$ represents its
``decay constant'' defined by $\langle0|j|h\rangle=f_hm_h$ (in the scalar case), $s_0$ denotes the onset of perturbative continuum
in terms of the energy squared $s$. Using a certain dispersion relation that relates~\eqref{12} and~\eqref{15} one finally derives
(with the help of~\eqref{13} and~\eqref{14}) an expression for $m_h$. In the case of scalar quark current $j=\bar{\psi}\psi$, the
final answer is~\cite{rry}
\begin{equation}
\label{16}
m_h^2\simeq M^2\frac{2a_0\left[1-\left(1+\frac{s_0}{M^2}+\frac{s_0^2}{2M^4}\right)e^{-s_0/M^2}\right]+\frac{\tilde{C}_3}{M^6}}
{a_0\left[1-\left(1+\frac{s_0}{M^2}\right)e^{-s_0/M^2}\right]+\frac{\tilde{C}_2}{M^4}-\frac{\tilde{C}_3}{M^6}},
\end{equation}
where $M$ is the Borel parameter in~\eqref{14} and $\tilde{C}_2$, $\tilde{C}_3$ are certain combinations of quark and gluon
condensate terms~\cite{rry}.

The hadron mass $m_h$ in~\eqref{16} depends on two parameters, $s_0$ and $M$. The Borel parameter $M$ must be large enough to suppress
the contributions from non-perturbative condensate terms in the OPE~\eqref{13}, otherwise the method does not work. The fixation of $s_0$
requires some guess on a ``natural'' value of continuum threshold in the hadron channel under consideration~\cite{rry}. On practice, $s_0$ is chosen
in a region where contributions from the higher resonances are negligible (but below the mass squared of the next resonance which should be
roughly estimated). Having fixed $s_0$ one finds the region of minimal sensitivity
to $M$ in~\eqref{16}, usually it corresponds to the minimum of function $m_h(M)$ in~\eqref{16}. The value of $m_h(M)$ in that stability region
gives the physical mass $m_h$.

In practical cases, the dependence of obtained value for $m_h$ on the values of condensates is rather weak because the dominance of perturbative
logarithm is a necessary condition for applicability of the method~\cite{svz1,svz2}. For the resonances heavier than the $\rho$-meson the minimum (or maximum)
of expressions like~\eqref{16} is very shallow, if it exists at all, and close to the asymptotics $M^2\rightarrow\infty$. This property was used
in~\cite{Afonin:2016hia} to derive simple expressions for the masses of scalar and axial-vector mesons. Taking in~\eqref{16} the limit $M^2\rightarrow\infty$ we
get for the mass of scalar meson~\cite{Afonin:2016hia}
\begin{equation}
\label{17}
m_h^2\simeq \frac23 s_0.
\end{equation}
Remarkably, the simple relation~\eqref{17} does not depend on the condensate terms but nevertheless it turns out to be a very good approximation
to~\eqref{16} within the accuracy of the method. For instance, the classical review~\cite{rry} reports the following numerical prediction for the mass
of scalar meson in~\eqref{16}: $m_h=1\pm0.03$~GeV, where the input $s_0\simeq1.5$~GeV$^2$ was used. This numerical result follows from~\eqref{17}
immediately.

Now let us move on to the Higgs. If the supposed strong BSM dynamics is similar to QCD, one may expect that the same method of spectral sum rules should work.
We do not know the ``BSM condensates'' but, fortunately, we may expect that the approximation~\eqref{17} is valid. Then the question arises what
value of ``perturbative threshold'' $s_0$ should we use? We suggest that a reasonable and natural candidate for $\sqrt{s_0}$ is
the upper bound on the Higgs mass. The tree-level perturbative unitarity leads to $m_h\lesssim 710$~GeV (see, e.g.,~\cite{Djouadi:2005gi} and references therein).
However, there is a known caveat to this estimate: The Higgs boson self-coupling, $\lambda=\frac{m_h^2}{2v^2}$, becomes strong near such large $m_h$,
so the radiative corrections are able to completely invalidate this estimate. One needs lattice simulations in which the non-perturbative effects
are automatically taken into account. Such simulations were performed by several groups for the triviality bound in the SM scalar sector~\cite{Heller:1993yv}
(see also the relevant discussions in~\cite{Djouadi:2005gi}). They reported the bounds: $m_h\lesssim620$~GeV~\cite{Bhanot:1988ua},
$m_h\lesssim640$~GeV~\cite{Kuti:1987nr}, and $m_h\lesssim660$~GeV~\cite{Hasenfratz:1988kr}.
Their average is $m_h\lesssim640\pm20$~GeV. Making use of this bound for $\sqrt{s_0}$ in~\eqref{17},
\begin{equation}
\label{17b}
\sqrt{s_0}\simeq 640\pm20~\text{GeV},
\end{equation}
we get the following estimate for the mass of second scalar in the Higgs sector,
\begin{equation}
\label{18}
m_{h'}\simeq 520\pm20~\text{GeV}.
\end{equation}

One might wonder where the standard Higgs boson $h$ is in this approach. If $h$ is some kind of pseudo-Goldstone boson in the hypothetical BSM sector, as is often assumed, then the situation with $h$ should be similar to the pion case in QCD sum rules: the perturbative logarithm in the OPE does not dominate, so the applicability condition of the method is violated~\cite{svz1,svz2,rry}. Thus, if $h'$ is not a radial excitation of $h$, then $h$ corresponds to a pole of another correlator, for which the method does not work.


\section{Extra Higgs from the NJL approach}

Below we first give a very brief review of the simplest NJL model invariant (in the massless limit $m_0=0$) under $U(1)_L\times U(1)_R$
chiral transformations. Our presentation of this model will follow the reviews~\cite{klevansky,Vogl:1991qt}.

Consider the following Lagrangian for a self-interacting fermion field $\psi$,
\begin{equation}
\label{1}
\mathcal{L}=\overline{\psi}\left(i\!\!\not\!\partial-m_0\right)\psi+G\left[(\overline{\psi}\psi)^2+(\overline{\psi}i\gamma_5\psi)^2\right].
\end{equation}
If a non-zero v.e.v. (condensate) $\langle\overline{\psi}\psi\rangle$ is formed due to some dynamics, the scalar part
of four-fermion interaction in~\eqref{1} can be linearized as
$(\overline{\psi}\psi)^2\simeq2\langle\overline{\psi}\psi\rangle\overline{\psi}\psi+\dots$.
In the Hartree (mean field) approximation, the Dirac equation for a single fermion, in the first approximation, is then
\begin{equation}
\label{2}
\left(i\!\!\not\!\partial-M\right)\psi=0,
\end{equation}
where
\begin{equation}
\label{3}
M=m_0-2G\langle\overline{\psi}\psi\rangle
\end{equation}
is a dynamically generated fermion mass. Eq.~\eqref{3} is called the gap equation by analogy with its non-relativistic realization
in the theory of superconductivity.

Within the mean field approximation, the condensate $\langle\overline{\psi}\psi\rangle$ is given by a one-point fermion loop
in which the fermion has the dynamical mass $M$,
\begin{equation}
\label{4}
\langle\overline{\psi}\psi\rangle=-\text{Tr}\int\frac{d^4\!p}{(2\pi)^4}\frac{i}{\not\!p-M+i\varepsilon}.
\end{equation}
Eq.~\eqref{3} represents thus a non-linear relation between $M$ and the coupling $G$. A non-trivial solution
of~\eqref{3} emerges when $G$ exceeds some critical value~\cite{klevansky}.

The integral in~\eqref{4} is divergent and requires a regularization. Since the model is nonrenormalizable,
it is defined not only by a Lagrangian but also by a chosen way for regulating divergent quantities, i.e.,
a specific regularization procedure finally determines a concrete model. A choice of regularization scheme
should be driven by expected physical properties. There are many schemes within which the NJL model was
considered in the literature: The noncovariant three-momentum cutoff and various covariant schemes such as
the four-momentum cutoff in Euclidean space, the regularization in proper time, and the Pauli-Villars (PV)
method~\cite{klevansky}. Only within the latter one the Lorentz and gauge invariance are maintained simultaneously.
These two fundamental symmetries are also preserved in the minimal subtraction scheme but this scheme does not
contain a physical cutoff which limits applicability of the effective model under consideration. In our case of
modeling a BSM physics, the PV-scheme looks thus the most attractive and it will be used in what follows.

The simplest NJL model~\eqref{1} describes a scalar and pseudoscalar fermion-antifermion excitations.
Below we will not be interested in the latter as we will not assume any chiral symmetry in the BSM strongly-coupled sector.
The mass of the scalar excitation in the model is given by the Nambu relation~\cite{klevansky},
\begin{equation}
\label{6}
m_\sigma\simeq 2M,
\end{equation}
which is exact in the limit $m_0=0$.
The condensate~\eqref{4} in the PV-regularization is given by~\cite{klevansky} (per light flavor)
\begin{equation}
\label{9}
\langle\overline{\psi}\psi\rangle=
-\frac{N_c}{4\pi^2}M\!\left[2\Lambda^2\ln\frac{1+2\Lambda^2/M^2}{1+\Lambda^2/M^2}-M^2\ln\frac{\left(1+\Lambda^2/M^2\right)^2}{1+2\Lambda^2/M^2}\right],
\end{equation}
where $\Lambda$ denotes the PV-cutoff. Given the phenomenological values for $\langle\overline{\psi}\psi\rangle$ and $\Lambda$ this relation
yields a numerical solution for $M$, hence, for $m_\sigma$ via~\eqref{6}.

Let us apply the relations above to the hypothetical BSM sector. By assumption, the electroweak symmetry breaking parameter, $v=246$~GeV, emerges
from some condensate $\langle\overline{\psi}\psi\rangle\equiv\langle0_\text{\tiny BSM}|\overline{\psi}\psi|0_\text{\tiny BSM}\rangle$ (averaging is understood
in the vacuum of a BSM strongly-coupled theory) and the fermion field $\psi$ is not yet specified --- it could be the $t$-quark or some new fermion.
The v.e.v. of SM Higgs field is $|\varphi_0|=v/\!\sqrt{2}=174$~GeV (it practically coincides with the mass of the $t$-quark within the experimental errors~\cite{pdg}).
We associate $|\varphi_0|$ with $\langle\overline{\psi}\psi\rangle$ in~\eqref{9}.

Now consider the following question: At which value of $N_c$ in~\eqref{9} we closely reproduce the estimate~\eqref{18} for $m_{h'}$ made in the previous Section if the
cutoff $\Lambda$ is set equal to the unitarity bound~\eqref{17b}? It turns out that this happens for $N_c=2$, for which Eq.~\eqref{9} gives
\begin{equation}
\label{11}
m_{h'} \simeq 480^{-50}_{+70}~\text{GeV}.
\end{equation}
For $N_c=1$ there is no solution, for $N_c=3$ we would have $m_{h'}\simeq280$~GeV and smaller values for larger $N_c$.

Thus we see that the two completely different approaches lead to a similar result for $m_{h'}$ if the BSM sector is described by a strongly coupled field theory
based on the $SU(2)$ gauge group. It is tempting to associate this group with $SU(2)_R$ --- the absent counterpart of $SU(2)_L$ gauge group of electroweak interactions.

\section{Discussions}

We have obtained a BSM scalar boson with the mass near 0.5~TeV using two very different approaches --- the spectral sum rules and a NJL-like model.
Both approaches contain a cutoff which was identified with the unitarity bound on the Higgs mass.

The cancelation of quadratic divergence in vacuum energy might provide a further, very heuristic but interesting argument in favor of possible existence of
extra Higgs-like boson near 0.5~TeV.
As was proposed long ago by Pauli~\cite{pauli} the vacuum energy density in the Universe must be almost equal to zero. Later Zeldovich~\cite{zeld}
imposed this condition to explain the observable vanishingly small value of the cosmological constant.

Let us remind the reader the essence of the idea. In the limit of free fields, the vacuum energy
density can be easily written by summing up the zero-point energies of free harmonic oscillators, $\pm\frac12\hbar\omega(k)$ (we will set $\hbar=1$),
with all possible three-momenta for all physical field components of mass $m_n$ up to a wave number cutoff $\Lambda$~\cite{zeld},
\begin{equation}
\label{19}
\rho_\text{vac}=\sum_n (-1)^{2S_n}g_n\frac12\,\int_0^\Lambda\frac{d^3k}{ (2\pi)^3}\,\sqrt{k^2+m_n^2},
\end{equation}
where the boson contributions are counted as positive and fermion contributions as negative. The degeneracy factor $g_n$
includes a spin factor ($2S_n + 1$ for massive particles and $2$ for massless ones), an additional factor of 2 when particle and antiparticle
are distinct, and also the factor of $N_c=3$ due to color (for example, $g_n=2\cdot 2\cdot3=12$ for quarks).
Integrating the relation~\eqref{19} and expanding the answer at $\Lambda\gg m_n$,
one arrives at the expression (see, e.g.,~\cite{Visser:2016mtr,Kamenshchik:2018ttr,Afonin:2022lst})
\begin{equation}
\label{20}
\rho_\text{vac}=\sum_n (-1)^{2S_n}g_n\left\{\frac{\Lambda^4}{16\pi^2}+\frac{\Lambda^2m_n^2}{16\pi^2}-\frac{m_n^4}{32\pi^2} \left[\,\ln\frac{\Lambda}{m_n}-\frac14+\ln2+O\left(\frac{m_n}{\Lambda}\right)^2\,\right]\right\}.
\end{equation}
Now we should impose the Pauli--Zeldovich condition,
\begin{equation}
\label{21}
\rho_\text{vac}=0.
\end{equation}

Obviously, taking the ultraviolet limit $\Lambda\rightarrow\infty$ in~\eqref{20} one obtains
three kinds of divergences which must be canceled. The compensation of the main quartic divergence,
\begin{equation}
\label{22}
\sum_n (-1)^{2S_n}g_n=0,
\end{equation}
was among early motivations for supersymmetry in particle physics. Later this cancelation became a source
of hypotheses about the particle content of dark matter (see, e.g,~\cite{Afonin:2022lst} and references therein).
One can show, however, that the appearance of quartic divergence on the level of free fields is an artifact of the use of non-covariant regularization~\cite{Akhmedov,Sirlin}.
In an interacting field theory, the quartic divergence appears even in covariant schemes of regularization~\cite{Sirlin}. Sometimes this observation is used in the
BSM phenomenology~\cite{Kamenshchik:2018ttr}.
We will assume that the BSM interactions are self-tuned so as to compensate the leading divergence in $\rho_\text{vac}$.

The cancelation of quadratic divergence in~\eqref{21} yields
\begin{equation}
\label{23}
\sum_n (-1)^{2S_n}g_n m_n^2=0.
\end{equation}
This condition depends on masses of fundamental particles which are determined by some unknown BSM dynamics. In this sense, the field interactions are already
taken into account in~\eqref{23}, at least partly. For this reason, one can assume that the interactions do not change much the condition~\eqref{23} with physical masses and that the
account for presumably small quantum corrections to~\eqref{23} should simultaneously provide cancelation of logarithmic divergence in~\eqref{20}. Then the condition~\eqref{23}
leads to an approximate relation between masses of fundamental particles. This relation was widely used in the literature as a guiding principle to estimate
masses of new particles in various BSM scenarios (a short review is given in~\cite{Afonin:2022lst}).

Consider now the contributions to~\eqref{23} from the SM particles. This contribution, of course, depends on the energy scale.
We will be interested in nullification of the vacuum energy density at typical EW energies.
The fermion sector is absolutely dominated by the top quark mass while the boson sector is
completely contributed by the masses of Higgs and weak gauge bosons. Taking into account the degeneracy factor, we get for~\eqref{23}
\begin{multline}
\sum_n (-1)^{2S_n}g_n m_n^2\approx m_h^2 + 3(2m_W^2 + m_Z^2)- 12  m_t^2 \approx\\
\big[125^2 + 3\!\cdot\!(2\!\cdot\!80^2 + 91^2) - 12\!\cdot\!173^2 \big] \text{GeV}^2 \approx - 530^2\, \text{GeV}^2.
\label{24}
\end{multline}
The simplest extension of the SM particle content satisfying the known EW constraints consists in adding a singlet scalar boson.
Thus, applying the ``Occam's razor'' we see immediately that the most natural way to satisfy the Pauli--Zeldovich condition~\eqref{21}
is to add to~\eqref{24} a new scalar boson $h'$,
\begin{equation}
\label{25}
\sum_n (-1)^{2S_n}g_n m_n^2\approx m_{h'}^2 + m_h^2 + 3(2m_W^2 + m_Z^2)- 12  m_t^2 \approx 0,
\end{equation}
with the mass of $m_{h'}\approx530$~GeV. This prediction is in agreement with the results of our previous estimates~\eqref{18} and~\eqref{11}.

From the experimental side, it is very unlikely that a peak corresponding to $h'$ can be directly detected if the Higgs boson is the only SM particle
with which $h'$ interacts.
The current experimental situation with the searches for additional Higgs-like bosons at the LHC is described in Ref.~\cite{Kundu:2022bpy}.
It appears that the strongest signal has been observed by the CMS and ATLAS Collaborations near 650~GeV, it was dubbed $H(650)$.
Tentatively, the corresponding scalar resonance has
the total width approximately $100$~GeV that would be typical for a SM Higgs-like particle of the mass around 650~GeV. On the other hand, some
peculiarities in the observed $H(650)$ decays suggest that the $H(650)$ may represent two distinct resonances~\cite{Kundu:2022bpy}.
The given observation is not confirmed, and even not officially recognized by the CMS and ATLAS Collaborations. On the other hand, presently this is the
only observation that might give a hope to speculate on a possible observable effect of $h'$. If the reality of $H(650)$ is confirmed in the future
then we could interpret the corresponding peak as an enhancement effect related with the kinematical $hh'$-threshold,
\begin{equation}
\label{26}
m_{H(650)}\simeq m_h + m_{h'}\simeq 125~\text{GeV} + 530~\text{GeV}.
\end{equation}

Our simple argument is based on a high degree of locality of EW interactions which should suppress the direct production of $h'$ in the reaction $hh\rightarrow h'$
washing out a possible peak around 0.5~TeV. However, if there is a vertex for the decay $h'\rightarrow hh$ then the reaction $h\rightarrow hh'$ for ultrarelativistic Higgs is
possible via the same vertex. The subsequent decay $h'\rightarrow hh$ entails the chain $h\rightarrow hh'\rightarrow hhh$ that contributes to
the quartic coupling $\lambda$ in the Higgs potential and should thereby lead to deviations of physical value of $\lambda$ from the perturbative
SM prediction. The given channel for the production of $h'$ should not be suppressed by very small effective size of the Higgs boson. Thus we
arrive at a qualitative physical picture for the dominant mechanism of $h'$-production: This is an associated $hh'$-production.
Consequently, $h'$ should be revealed not at the c.m.s. energy $E\approx m_{h'}$ but at $E\approx m_h + m_{h'}$.

It is interesting to note that our prediction of the second
Higgs from the interpretation of cutoff as an approximate unitarity bound may somewhat resemble the appearance of dynamical resonances near the scale of
unitarity violation in effective field theories within the framework of inverse amplitude method, an application of this method to description
of a second Higgs-like scalar boson is considered in Ref.~\cite{Asiain:2023myt}. This analogy can be developed further: We implied
that our models deal with a description of strongly-coupled BSM physics but this interpretation is not unique. The $h'$-boson, if it exists,
may turn out to be an $hh$-resonance in a similar sense as the $\sigma$-meson is observed as a $\pi\pi$-resonance. In relativistic field theories,
there is no analytical understanding of appearance of such resonances. The existence of $\sigma$-meson was deduced from the phenomenology of nuclear
and particle physics --- analytically it is totally not obvious why the pion exchanges responsible for nuclear forces and low-energy $\pi\pi$-scattering
are largely contributed by a highly correlated exchange of $\pi\pi$ pair with quantum numbers of the scalar isoscalar meson. And it is not
clear why correlation in the exchanging pair is so strong that this $\pi\pi$ pair can be well described as a new hadronic particle, i.e., as a new
hadronic state which emerges dynamically and reveals all the properties of independent particle that contributes also into other reactions with the pions
(pion-nucleon scattering, decays of heavy mesons, etc.). Thus, if the $h'$ is indeed a complete electroweak
analogue of the $\sigma$ (i.e., represents a dynamical $hh$-resonance\footnote{This assumption alone could lead to the prediction $m_{h'}\simeq 0.5$~TeV:
If the underlying dynamics is, loosely speaking, in the same universality class as QCD, then the approximate numerical relation $m_\sigma\simeq 4m_\pi$
may be used as a guide to predict $m_{h'}\simeq 4m_h$.}) then it hardly can shed light on the BSM sector. On the other hand, the
formation of the $h'$ should then affect the $hh$ production in the future high-luminosity LHC run causing a distortion from the perturbative SM
prediction. For instance, we would expect an enhancement (due to a resonance) of measured value of trilinear Higgs coupling while the scenarios
with new fundamental scalar(s) predict the opposite effect --- an observable deficit in this value~\cite{MartinLozano:2015vtq,No:2013wsa,Chen:2014ask}.
Thus we have a possible experimental test to distinguish these two scenarios in the future LHC experiments.

\section{Conclusions}

Assuming the existence of strongly-coupled BSM sector we obtained a heavy Higgs-like scalar boson with the mass $m_{h'}\simeq0.5$~TeV
within the framework of two substantially different model approaches to effective description of strongly-coupled gauge theories ---
the spectral sum rules and the NJL model in the scalar sector. The first model is a particular case of approaches in which composite states are represented by their
interpolating fermion currents. The second one is a representative of a large class of approaches in which composite states are
treated in terms of constituent fermions. Both models contain a four-momentum cutoff, and our result follows if this cutoff is identified with the
unitarity bound on the Higgs mass $m_h$. The results of two approaches are close if the underlying BSM sector is described by some field theory based
on the $SU(2)$ gauge group.

We also proposed some additional qualitative arguments in favor of the existence of an isosinglet heavy scalar boson near 0.5~TeV.

The present study continued the discussions on the value of mass of a possible second Higgs-like boson which were started in our paper~\cite{plb},
where, making use of holographic approach to modeling BSM physics, its mass was also estimated near 0.5~TeV.

Taken separately, all the arguments mentioned above are rather heuristic. However, the observation that they independently point to the same result is intriguing.
We hope that the agreement found between several different approaches will encourage further research in the search for the hypothetical second Higgs boson.

\bigskip

\underline{\bf Conflict of Interest}: The author declares that he has no conflicts of interest.

\underline{\bf Funding}: This work was funded from the budget of the Saint Petersburg State University. No additional grants have been received to conduct or direct this particular study.



\end{document}